\documentclass[useAMS,usenatbib]{mn2e}
\usepackage{graphicx}
\usepackage{txfonts}
\usepackage{natbib}
\newcommand{\xte}{{\it RXTE}}

\def\gsim{\mathrel{\hbox{\rlap{\hbox{\lower4pt\hbox{$\sim$}}}\hbox{$>$}}}}
\def\lsim{\mathrel{\hbox{\rlap{\hbox{\lower4pt\hbox{$\sim$}}}\hbox{$<$}}}}

\begin{document}
   \title[]{Quasi-periodic Oscillations in XTE J1550-564: The rms-flux relation}

   \author[L.M. Heil, S. Vaughan, P.Uttley]{L.M. Heil$^{1}$, S. Vaughan$^{1}$, P. Uttley$^{2}$\\
   $^{1}$ X-Ray and Observational Astronomy Group, University of
   Leicester,  Leicester. LE1 7RH, U.K.\\
   $^{2}$ School of Physics and Astronomy, University of Southampton, Southampton, SO17 1BJ, U.K.}

   \date{Draft \today}

   \pagerange{\pageref{firstpage}--\pageref{lastpage}} \pubyear{2002}

   \maketitle
   
   \label{firstpage}

   \begin{abstract}
	We present an analysis of the short timescale variations in the properties of the strong (type ``C'') quasi-periodic oscillation observed in XTE J1550-564 during its 1998 outburst. In particular, the QPO shows a correlation between absolute rms amplitude and mean source flux over timescales shorter than $\sim 3$ ksec. A linear rms-flux relation has been observed to be a common property of broad-band noise but here we report the first detection of rms-flux dependence in a QPO. The gradient of the rms-flux relation is correlated with the QPO peak frequency: from a strong positive correlation when the QPO peak frequency is below $\sim 4$ Hz, through no correlation, to a strong negative correlation when the peak frequency is above $6$ Hz. This is the first time a negative short term rms-flux relation has been observed in any component of the power spectrum. Previous work on both the broad-band noise and QPOs in a range of sources have suggested the presence of a filter reducing the amplitude of QPOs with increasing frequency. We attempt to remove the possible effects of this filter and find that the previously negative rms-flux relations above $\sim 5$ Hz become constant.

   \end{abstract}

 

\section{Introduction}
\label{sect:intro}
  Accreting black holes show variations over a broad range of timescales. In the well-studied Galactic Black hole binaries (BHBs) their power spectra may include broad-band noise, band-limited noise (BLN) and quasi-periodic oscillations (QPOs), distinguished by their relative frequency widths \cite[see e.g.][for reviews]{vanderKlis06, Remillard06}. The exact origin of the variability and its characteristic frequencies remain elusive, although strong correlations between frequencies of different power-spectral components have been observed as well as correlations between frequencies and spectral parameters \citep[see e.g.][]{Klein-Wolt08, Belloni02, Remillard02}. 

QPOs are observed over a range of frequencies in BHBs and NS, in BHBs they are commonly described as either low frequency $<$ 50 Hz or high frequency $>$100 Hz \citep{vanderKlis06}. Low frequency QPOs are often observed with further harmonics and sub harmonics and their formation mechanism is unclear. The three main types are commonly classified as A, B or C \citep{Wijnands99, Homan01, Belloni02, Casella04, Casella05}, with types A and B observed in the soft intermediate states, and displaying different harmonic and phase lag behaviours, and the type Cs observed in hard states (i.e low-hard and hard intermediate) with strong broad-band noise. The frequency of type C QPOs is known to correlate with spectral properties of the source, the photon index, disc inner radius, disc temperature and both power law and disc fluxes \citep[see e.g.][]{Remillard02, Muno99} although there is as yet no consensus on the physical origin of QPOs \citep{vanderKlis06, Remillard06, Done07}. 

One of the simplest and most stable relationships observed between variability properties is the rms-flux relation, which connects the rms amplitude of variations to the mean flux level by a positive linear correlation that appears to operate over a wide range of timescales where the power spectra remain similar \citep{Uttley01, Gleissner04}. This relation is inherent to the short-term variability of sources with an otherwise-stationary (or close to stationary) power-spectral shape. Longer-term changes in rms can be caused by changes in power-spectral shape which correlate with the spectral evolution of the source within and between the different spectral states, but these are different to the rms-flux relation which appears to be a fundamental aspect of the variability process itself. The linear rms-flux relation was initially observed in the BHB Cygnus X-1, to date the same linear rms-flux relation has been observed in Active Galactic Nuclei \citep{Uttley01, Vaughan03, Gaskell04, Uttley05}, one neutron star X-ray binary \citep{Uttley01, Uttley04} and one Ultra-Luminous X-ray source (ULX) \citep{Heil10}, it has also recently been observed over long timescales in the BHB GX 339-4 \citep{Munoz10}. This relation is observed in the noise components of the power spectrum, and in the $401$ Hz pulsation of the neutron star system SAX J1808.4-3658 \citep{Uttley04}, but to date has not been studied for QPOs.

Arguably the most promising explanation for the linear rms-flux relation observed for the broad-band noise of accreting systems is the `propagating fluctuation' model (see discussion and references in \cite{Uttley01, Gleissner04, Uttley04, Uttley05}). In this general model long timescale fluctuations in the accretion rate originate from large radii in an accretion disc (perhaps as random variations in viscosity) and propagate inwards, modulating shorter timescale variations originating at smaller radii \citep[e.g.][]{Lyubarskii97, King04, Arevalo06}, and eventually modulating the accretion rate in the innermost, X-ray producing regions. This simple scheme naturally explains X-ray emission that varies on a wide range of timescales, and a linear rms-flux relation due to the multiplicative coupling of variations on all timescales. It is not clear how QPOs fit into this scheme.

Secular ($\sim$ 1 day) variations in the QPO properties (rms and frequency) are
known to be correlated with the (energy spectral) evolution of black hole transient outbursts
(i.e. changes in the physical state of the system). But this overall
evolution of the outburst properties tends to follow distinct patterns, such
as the hysteresis curve on the hardness-intensity diagram \cite[HID; see e.g.][]{
Belloni05, Belloni10}. As previously discussed, the within-observation
variations (timescales $<$ 3ks studied here) are variations in an
almost stationary process (presumably random fluctuations within the
accretion flow), the average properties of which evolve through outburst
\citep[see][and Figure \ref{fig:QPOrmsflux}]{Munoz10}. It is these relatively small and 
flux-correlated deviations from stationarity, and how they relate 
to other source properties, that are the subject of this paper.

We focus on the type "C" QPOs observed at the beginning of the the 1998 outburst of XTE J1550-564. The long-term properties of this source have been studied extensively by \cite{Cui99}, \cite{Wijnands99}, \cite{Homan01}, \cite{Remillard02}, \cite{Sobczak00}, \cite{Reilly01}, \cite{Chakrabarti09}, and \cite{Rao10}. \cite{Sobczak00} compared the changes in amplitude and frequency for the QPOs over the course of the outburst to the energy spectral components in both XTE J1550-564 and the 1996-97 outburst of GRO J1655-40. They found that the QPO frequency appears to be linked to both the disc flux and the strength of the power law. They also observe that for the outburst of XTE J1550-564, the amplitude of the QPO is closely linked to the QPO frequency: it rises until around 3 Hz but appears to then fall off above this point. \cite{Rao10} split each observation into 128s segments and demonstrated that over time the frequency of the fundamental QPO component can be linked to the disc count rate suggesting that there is some correlation between frequency, flux and QPO amplitude over short timescales. We further examine the short term variations in these parameters for the QPO, concentrating on the rms-flux relation within each observation and its relation to the peak frequency.

\section{Observations}

We use observations taken by the PCA detector on \xte~ from the 1998 outburst of XTE J1550-564 in programs P30188 and P30191, covering the first 46 days of the outburst. The data modes used all had 8 bit wide counters and are not likely to suffer the buffer overflow problems which can distort the rms-flux relation away from a linear relationship \citep[seen in][Appendix A]{Gleissner04}. During the first half of the outburst the source reached a very bright intermediate state where the strong (``type C") QPO shows high rms amplitudes, this means that they are ideal candidates to test the behaviour of the rms-flux relation for QPOs and how this relates to the continuum.

Binned mode data has been used for this analysis. There are a few variations of this data mode used over the outburst, so the energy ranges for some of the observations differ. The ranges have been chosen so that they are as consistent across the sample as possible, although slight variations remain between observations. The energy range chosen was as close to 4-14 keV as possible, and the difference between energy bands is no greater than 0.5 keV.


\section{Data Analysis}
\label{sect:data}

In order to measure the rms-flux relation each continuous light curve is divided into non-overlapping 3 s intervals. Due to the lower frequency limit imposed by this segment size we only use observations from this sample where the primary type C QPO is above 1 Hz. In practice this includes most of the observations where the QPO is detected, only excluding six, leaving 48 observations to be processed. From each 3 s segment the power spectrum was estimated in absolute normalisation using standard techniques, and the background subtracted mean count rate of the interval, corrected to 1 PCU detector, was recorded. The power spectral estimates were then averaged in non-overlapping flux bins, with at least 20 intervals contributing to each bin, to provide well-determined power spectra as a function of flux. These power spectra were then fitted in {\sc xspec 12} \citep{Arnaud96}. 

The models discussed in \cite{Rao10} have been applied, where multiple Lorentzians are used to describe the fit \citep[see figure 1 of][for representative fitted power spectra]{Rao10}. In all observations the model used consisted of four Lorentzians: Two of these were fitted to the broad band noise, they are referred to in \cite{Rao10} as $L_{ft}$ and $L_{pn}$ ($ft$ stands for flat topped noise at low frequencies $<$ 1 Hz and $pn$ for peaked noise, generally seen at frequencies close to, or just above, the QPO harmonic). These two Lorentzians had Q-factors less than 2 ($Q = \nu/\Delta\nu$ where $\nu$ is the centroid frequency and $\Delta\nu$ the full width at half maximum). Following \cite{Rao10} the primary QPO component is defined as the strongest feature in each observation. Due to the limited frequency range (0.33 - 100 Hz) within the flux-binned power spectra it is rarely possible to resolve the fundamental and both the harmonic and subharmonic simultaneously. For this reason the two further Lorentzians described the primary QPO component and either the harmonic or the sub-harmonic, depending on which was most visible. These Lorentzians are referred to as $L_{F}$ and either $L_{s}$ or $L_{h}$. As in \cite{Rao10} the Poisson noise component was not removed from the power spectra, but is included as an additional constant in the model. This model provided an adequate fit to all observations within the sample, the reduced chi-squared varied from 0.75 to 1.41, with the worst fit having a chi-squared of 354.1 with 179 degrees of freedom. \cite{Rao10} indicated the need for a further high frequency broad-band component in some observations within their sample. All of these observations are ones where the QPO is below 1 Hz, therefore this was not necessary for the fitting of any observations used here. For each fit the Lorentzian properties were recorded, namely the total rms (in absolute units, i.e ct/s/PCU), the peak frequency ($\nu_{F}$) and the width. See \cite{Pottschmidt03} section 3 for appropriate formulae. The rms value is corrected to represent the power in the QPO over only positive frequencies. This normalisation is thus the integrated power in the QPO ($R^{2}_{F}$) and the measured rms within the component is therefore $\sigma=\sqrt{R^{2}_{F}}$. 

\section{Results}

\begin{figure}
\begin{center}
   \includegraphics[width=9.0 cm, angle=0.0]{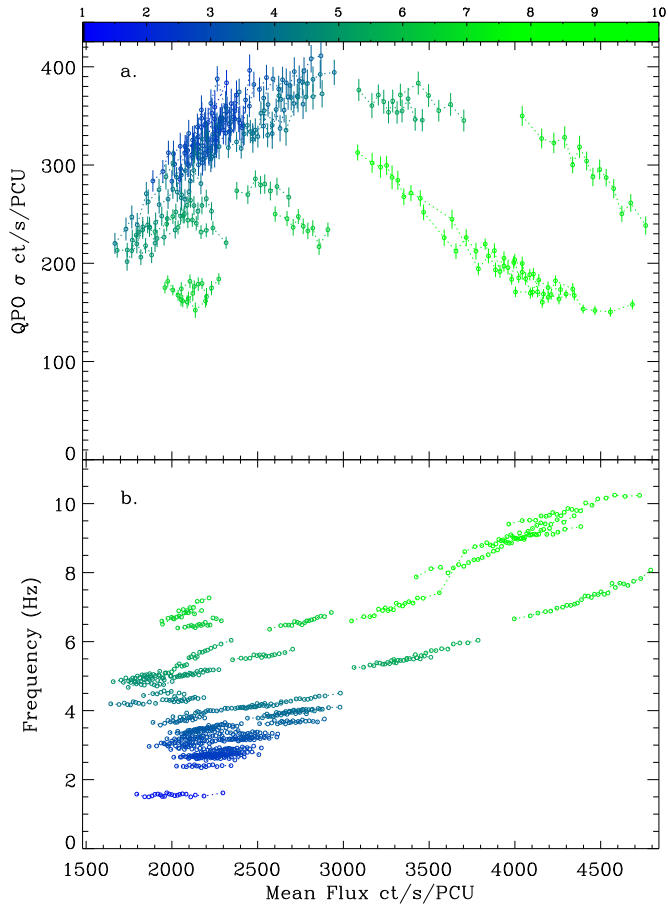}
\end{center}
\caption{\emph{a.} Rms-flux relations for the fundamental QPOs in each observation within the sample. Points within observations are joined by dotted lines. Some observations with rms values close to 350 ct/s/PCU and mean count rates between 2000-2500 ct/s have been removed for clarity. \emph{b.} Measured fundamental QPO frequency for each of the flux bins used for the QPO rms-flux relations. The steeper change in frequency with flux at high frequencies is clearly visible. For both \emph{a.} and \emph{b.} the colour bar describes the average frequency of the fundamental QPO component.}
\label{fig:QPOrmsflux}
\end{figure}

\begin{figure}
\begin{center}
   \includegraphics[width=6.0 cm, angle=90]{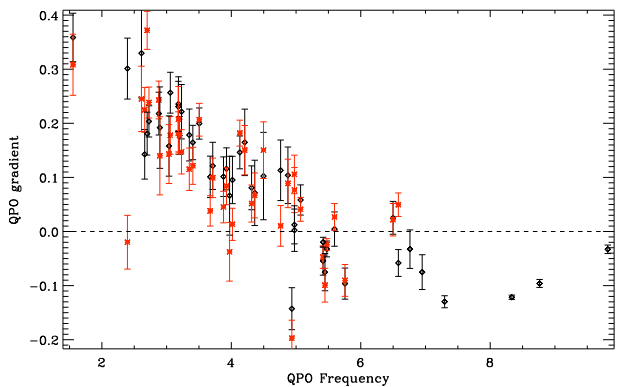}
\end{center}
\caption{Gradient ($k$) of the rms-flux relation against frequency of the fundamental QPO measured in both the fundamental (\emph{diamonds - black}) and the harmonic (\emph{crosses - red}). Both clearly follow a very similar pattern.}
\label{fig:gradharm}
\end{figure}

Fig. \ref{fig:QPOrmsflux} shows the dependence of the QPO amplitude (rms) and frequency on flux (count rate) both within and between observations (typically one day apart). Clearly the QPO frequency is varying on short timescales and is well correlated with the flux as mentioned previously by \cite{Rao10}, \cite{Sobczak00} and \cite{Remillard02}, the gradient of this relation appears to increase with average frequency. The rms shows a more complex relation with the flux: when the QPO is below 4 Hz it follows the positive linear rms-flux relation observed in the broad-band noise, between  4.0-5.5 Hz the rms becomes constant with flux, above 5.5 Hz the rms-flux relation is negative. This is the first time a short-term negative rms-flux relation has been observed. 
The average QPO frequency for an observation is not monotonically related to either the mean rms or flux over long timescales, even though clear relationships between these variables are found within each observation (see Figure 1). This illustrates the different effects of the long term evolution of the system vs. short term random variations in the accretion flow.
 
Fitting the rms-flux relations  for each observation with a linear function gives a reasonable measure of the gradient ($k_F$) which is clearly anti-correlated with the average QPO frequency ($\nu_F$ as seen in Figure \ref{fig:gradharm}. The point where the gradient reverses from positive to negative occurs around 5.5 Hz. The gradient of the rms-flux relation for the harmonic, $k_h$, changes simultaneously with that of the fundamental, $k_F$. Figure 2 shows $k_h$ becomes negative when $k_F$ does, i.e. around $\nu_F > 5.5$ Hz (where $\nu_h > 11$ Hz) (Figure \ref{fig:gradharm}, red points). 
We have been unable to asses the flux dependence of the high frequency power spectral continuum due to uncertainties in subtracting the power of the QPO and harmonics.

\section{Discussion}

\subsection{Summary}

We have shown that the rms amplitude ($\sigma_F$) and peak frequency
($\nu_F$) of the ``type C" QPO of the microquasar XTE J1550-564 (observed
during the 1998 outburst) both vary strongly on short timescales ($3-3000$
s) and are correlated with the source flux. The rms-flux relation for the
QPO depends on the peak frequency of the QPO: when $\nu_F \sim 1- 5$ Hz
there is a positive, linear correlation between the QPO rms and flux, at
around $\nu_F \sim 5$ Hz the rms-flux relation becomes flat (rms independent
of flux), and at higher frequencies the rms and flux becomes negatively
correlated. For all values of $\nu_F$ examined ($\sim 1.5 - 10$ Hz) the
rms-flux relation is approximately linear, and the gradient itself appears
to be monotonically related to the QPO peak frequency. Above $\sim 3$ Hz $\nu_F$ also increases with flux in an approximately linear manner, below this point it remains constant. The correlations
between the flux and the QPO rms and peak frequency on these short
timescales are distinct from the long term evolution of the QPO during the
outburst that is apparent between observations separated typically by $\sim
1$ day.

Within the framework of the ``propagating fluctuation" model discussed in
the Introduction, it is reasonable to imagine the QPOs as an enhancement in
the variability amplitude at certain frequencies (radii) but still coupled
to the broad-band spectrum of variations. This would naturally lead to a
linear, positive correlation between long term flux (driven by variations at
frequencies lower than the QPO) and QPO rms amplitude. It is less obvious
what causes the frequency-dependent changes in the rms-flux behaviour of the
QPO. We explore some possibilities and implications below.

\subsection{The effect of a Frequency Dependent Filter}

Using these observations \cite{Sobczak00} showed that on long timescales the
fractional rms of the QPO rises with frequency until around 3 Hz, above this
it falls off steeply (see their figure 6). \cite{Pottschmidt03} and
\cite{Axelsson05} showed that the broad Lorentzian components of
the power spectrum of Cygnus X-1 also show a sharp drop in amplitude above 3-5 Hz
\citep[see figure 6 of][]{Pottschmidt03}. Similar correlations between QPO
amplitude and frequency have been observed in the BHBs H1743-322, GRO
J1650-500, XTE J1655-40 and GRS 1915+105 \citep{McClintock09, Kalemci06,
Kalemci03, Debnath09, Muno99}.

\begin{figure}
\begin{center}
   \includegraphics[width=8.0 cm, angle=0]{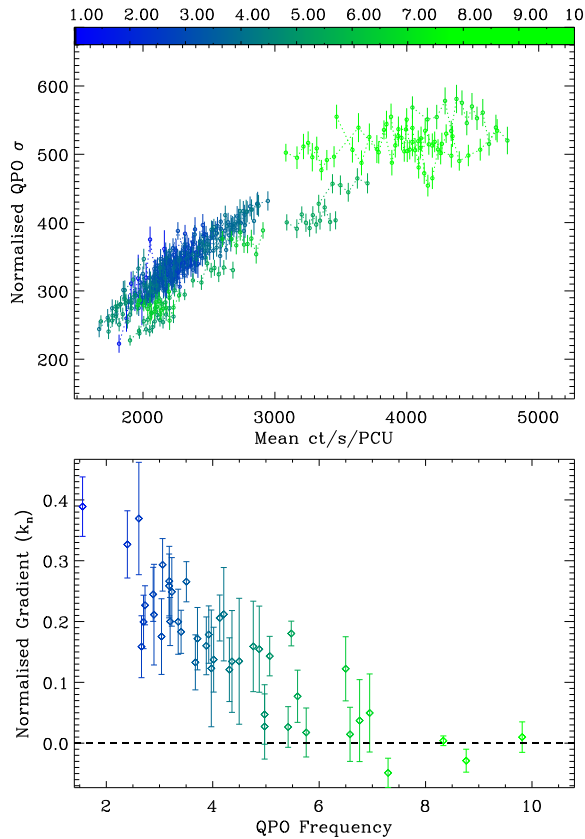}
\end{center}
\caption{\emph{a)} Rms-flux relations for the QPO following the removal of a filter reducing the amplitude of the QPO with increasing frequency. \emph{b)} Gradients found from fitting the above rms-flux relation with a linear function. Frequency dependence of the rms-flux relation is still visible, but the gradients no longer become negative.}
\label{fig:unfiltrms}
\end{figure}

One explanation for the attenuation of these components as they move to
higher frequencies is in terms of a low pass filter acting to suppress
variability above $\sim few$ Hz \citep{Done07, Gierlinski08}. Proposed physical mechanisms for the suppression of fast
 variations include the damping of accretion rate
 perturbations faster than the viscous timescale in the
 inner disc \citep{Psaltis00, Done07} and scattering of
 emitted X-rays by an optically thick outflow created at
 high luminosity \cite{Titarchuk07}. Such a
filter would affect the observed rms-flux behaviour for QPOs that move
in frequency above $\sim 3-5$ Hz. Is the change in $k_F$ simply the result of a QPO frequency that increases with flux on short timescales, combined with the rms being more strongly suppressed (by the action of the filter) at higher frequencies and thus fluxes? In order to investigate this possibility we modelled
the decay of the QPO (fractional) rms with frequency in terms of a doubly
broken power law, and use this to ``recover'' the unfiltered QPO rms.

A doubly broken power law over 1.5-10 Hz, with indices of $\alpha$ = -0.1, -0.5 and -1.7 and breaking at 3.4 Hz and 5.1 Hz respectively, can be used to describe the trend in the long term rms-frequency behaviour of the QPO. We note that this is
likely an over-estimate of the effect of a filter, since it predicts
filtering of the power spectrum by a power law with with index $-3.4$ above
$5$ Hz; the observed power spectrum is usually flatter. Assuming this
function describes the filter it is possible to ``correct" each data point
in Figure \ref{fig:QPOrmsflux} for the rms-reducing effect of the
frequency-dependent filter.

The rms-flux relations using the ``filter-corrected'' amplitudes are shown
in Figure \ref{fig:unfiltrms}. There are two significant differences between
this and the raw data shown in Figure \ref{fig:QPOrmsflux}a: at higher
frequencies the rms-flux gradient tends to zero above $\sim 5$ Hz, rather than
becoming negative, and the short timescale rms-flux data (within each
observation) now follow a similar curve to the longer timescale (between
observation) changes in rms and flux. Significantly for the present
discussion, the rms-flux relation still shows a monotonic dependence of
gradient on the QPO frequency. A flattening to zero gradient might indicate
a ceiling on the QPO strength (in absolute units) which effectively prevents
the rms increasing past $\sim 600$ ct s$^{-1}$ when $\nu_F > 5$ Hz.

\subsection{Physical implications of QPO behaviour}

Although we still lack a complete physical picture for the origin of QPOs
\cite[see e.g.][]{vanderKlis06}, recently proposed ideas include
Lense-Thirring precession of an extended, hot flow \citep[see
e.g.][]{Ingram09} and a magnetohydrodynamic dynamo cycle \citep{Oneill10}.
However, it is not clear what, if any, predictions such models make for the
rms-flux relation of the QPO. One would hope that as these and similar
models are developed and explored they will yield specific predictions about
the coupling of QPO amplitude and the lower frequency noise variations that
can then be compared against the observed behaviour.

We have shown there is a clear change in the flux-dependence of the QPO that
is itself linked to the QPO frequency, and that peak frequencies of $\nu_F
\approx 5-6$ Hz appear to stand out as marking where the rms-flux gradient
becomes zero. (This seems quite robust to the possible effects of a
frequency-dependent filter.) This frequency range is already noteworthy as
the location of the type B QPOs in this outburst and others. Given this
coincidence it seems reasonable to speculate that 5-6 Hz may represent a
relatively stable characteristic frequency of the system (in contrast to the
variable peak frequency of type C QPOs). Replication of these results for
other sources and outbursts is necessary to establish whether a fixed
characteristic frequency can be identified in this way.

\subsection{Correlations with the energy spectra}

It is well known that the properties of the type C QPOs in BHB outbursts
(notably $\nu_F$) are strongly correlated with the various energy spectral
parameters over long timescales, i.e. between observations \citep[see
e.g.][for discussion of the 1998 outburst of XTE J1550-564]{Sobczak00,
Remillard02, Rao10}. Using the spectral parameters obtained by both
\cite{Sobczak00b} and \cite{Dunn10} we find strong correlations between the
energy spectrum (e.g. power law index, disc flux, etc.) and the gradient $k_F$
of the rms-flux relation for the QPO, as would be expected given that all
these variables are strongly correlated with $\nu_F$. At the present time
the physical basis of these relations remains unclear.

\cite{Dunn10} noted that this outburst of XTE J1550-564 is unusual in that
it reaches a particularly high intermediate state, at the peak of the
outburst they find that the luminosity is close to L$_{Edd}$. It is during
these high flux observations, where the QPO reaches its highest frequency,
that the loss in positive correlation between rms and flux is most clearly
seen. Further investigations into the effect of high luminosities, and their associated states, on the
short-timescale behaviour of low frequency QPOs in other outbursts and
sources would be necessary to fully understand any link.

\section*{acknowledgements}

LMH acknowledges support from an STFC studentship. PU is supported by an STFC Advanced Fellowship, and funding from the European Community's Seventh Framework Programme (FP7/2007-2013) under grant agreement number ITN 215215 "Black Hole Universe". This research has made use of data obtained from the High Energy Astrophysics Science Archive Research Center (HEASARC), provided by NASA's Goddard Space Flight Center. We would like to thank the reviewer for their helpful comments and suggestions.

\bibliographystyle{mn2e}
\bibliography{qporms_v2.bib}

\bsp

\label{lastpage}

\end{document}